\documentclass{ws-mpla}
\usepackage[super]{cite}
\usepackage{graphicx}
\usepackage{url}
\usepackage{cancel}
\usepackage{hyperref}
\begin{document}

\markboth{S. Caudill, S. Kandhasamy, C. Lazzaro, A. Matas, M. Sieniawska, \& A. Stuver}
{Gravitational-wave searches in the era of Advanced LIGO and Virgo}

\catchline{}{}{}{}{}

\title{Gravitational-wave searches in the era of Advanced LIGO and Virgo}

\author{SARAH CAUDILL}
\address{Institute for Gravitational and Subatomic Physics (GRASP), Utrecht University, Princetonplein~1, 3584 CC, Utrecht, The Netherlands\\
Nikhef, Science Park 105, 1098 XG, Amsterdam, The Netherlands \\
s.e.caudill@uu.nl
}

\author{SHIVARAJ KANDHASAMY}
\address{Inter University Center for Astronomy and Astrophysics,
Pune,\\ Maharashtra - 411 007, India\\
shivaraj@iucaa.in
}

\author{CLAUDIA LAZZARO}
\address{Dipartimento di Fisica ed Astronomia, 
Università degli Studi di Padova,\\
Via G. Marzolo, 8 - 35131 Padova, Italy\\
INFN, Sezione di Padova, 
I-35131 Padova, 
Italy\\
claudia.lazzaro@unipd.it
}

\author{ANDREW MATAS}
\address{Max Planck Institute for Gravitational Physics (Albert Einstein Institute),\\
D-14476 Potsdam, Germany\\
andrew.matas@aei.mpg.de
}

\author{MAGDALENA SIENIAWSKA}
\address{Centre for Cosmology, Particle Physics and Phenomenology (CP3),\\
Universit\'{e} catholique de Louvain
Chemin du Cyclotron,\\ 2B-1348 Louvain-la-Neuve,
Belgium\\
magdalena.sieniawska@uclouvain.be
}

\author{\footnotesize AMBER L. STUVER\footnote{Corresponding author.}}
\address{Department of Physics, Villanova University, Villanova, PA 19085,
United States\\
\href{mailto:amber.stuver@villanova.edu}{amber.stuver@villanova.edu}
}

\maketitle

\pub{Received 29 June 2021}{Accepted 7 July 2021}

\newpage

\begin{abstract}
The field of gravitational-wave astronomy has been opened up by gravitational-wave observations made with interferometric detectors.  This review surveys the current state-of-the-art in gravitational-wave detectors and data analysis methods currently used by the Laser Interferometer Gravitational-Wave Observatory in the United States and the Virgo Observatory in Italy.  These analysis methods will also be used in the recently completed KAGRA Observatory in Japan.  Data analysis algorithms are developed to target one of four classes of gravitational waves.  Short duration, transient sources include compact binary coalescences, and burst sources originating from poorly modelled or unanticipated sources.  Long duration sources include sources which emit continuous signals of consistent frequency, and many unresolved sources forming a stochastic background.  A description of potential sources and the search for gravitational waves from each of these classes are detailed.

\end{abstract}


\section{Introduction}	

Gravitational waves carry information about the changing distribution of matter and detecting them provides humans with another medium to observe the Universe.  Specifically, sources of gravitational waves involve mass that is accelerating in a spherically asymmetric manner.  These sources are typically categorized into four classes: 
\begin{itemize}
    \item Compact binary coalescences (CBC) are binary sources of compact objects (black holes, neutron stars, white dwarfs) whose orbits shrink until the objects merge, q.v. section \ref{sec:cbc}.
    \item Burst gravitational waves originate from sources that are not well modeled (e.g. core collapse supernovae) or are unanticipated, q.v. \ref{sec:bursts}.
    \item Continuous gravitational waves have a consistent frequency over long periods of time originating from rapidly rotating objects, such as a neutron star with a surface defect, q.v. \ref{sec:continuous}.
    \item Stochastic gravitational waves are the unresolved gravitational waves from many different sources, such as primordial gravitational waves from the Big Bang, q.v. \ref{sec:stochastic}.
\end{itemize}

The field of multimessenger astrophysics consists of the observation of the same astronomical event in multiple media such as electromagnetic (EM) radiation, gravitational waves, or neutrinos.  The sources for both the CBC and burst searches are expected to be strong emitters of gravitational waves and potentially EM radiation.  LIGO and Virgo have a long history of performing follow-up observations of events observed in EM and neutrinos (e.g. Ref.~\refcite{Abbott:2008zzb,Abadie:2010uf,Briggs:2012ce,Abbott:2016cjt,Authors:2019fue}).  Now that CBC and burst searches can be  performed with low latency, the gravitational-wave community is issuing alerts of candidate gravitational-wave events. The application of gravitational-wave searches to multimessenger astrophysics is detailed in section \ref{sec:mma}.

A passing gravitational wave is transverse and will create alternating compression and expansion at orthogonal angles.  This creates a measurable strain ($h$) defined as a change in length ($\Delta L$) caused by the compression or expansion divided by the nominal length ($L$), $h = \frac{\Delta L}{L}$, in freely falling masses.  Two polarizations are permitted in standard general relativity: the plus-polarization, $+$, and the cross-polarization, $\times$, which is the plus-polarization rotated by $45^\circ$~\cite{Stuver2019, SchutzEnvelope, 300yrs, Maggiore2007}. 

Current methods of seeking gravitational waves include ground-based interferometric detectors and pulsar timing arrays (PTA).  Interferometers use lasers to directly measure the strain between suspended mirrors caused by a passing gravitational wave.  Current interferometric detectors have sensitive bandwidths between tens of Hz to several kHz and include:

\begin{itemize}
    \item \emph{LIGO}: The Laser Interferometer Gravitational-Wave Observatory (LIGO) is composed of two detectors operated by the United States located in Livingston, Louisiana, and Hanford, Washington. Each detector has two 4-km long arms.~\cite{TheLIGOScientific:2014jea}
    \item \emph{Virgo}: The Virgo detector located outside Pisa, Italy, has 3-km long arms.~\cite{aVirgo2014}
    \item \emph{KAGRA}: \underline{Ka}mioka \underline{Gra}vitational Wave-Detector (KAGRA) is a cryogenic detector with 3-km long arms constructed underground in the Kamioka Observatory in Japan.~\cite{aKAGRA}
    \item \emph{GEO600}: The GEO600 detector is located in Hannover, Germany with 600-m long arms and is a test bed for advanced detector technologies.~\cite{GEO2015}
\end{itemize}
    
Pulsar timing arrays (PTA) study deviations in predicted pulse arrival times on the order of tens of nanoseconds from several pulsars to measure correlations that are dependent on the changing curvature of space-time (gravitational waves) and not the pulsars themselves.  PTAs are sensitive to gravitational waves with nanohertz frequencies. The \emph{International Pulsar Timing Array (IPTA)}~\cite{IPTA2013} is made up of three collaborations: the North American Nanohertz Observatory for Gravitational Waves (NANOGrav)~\cite{NANOGrav2019}, the European Pulsar Timing Array (EPTA)~\cite{EPTA2013}, and the Australian-based Parkes Pulsar Timing Array (PPTA)~\cite{PPTA2013}.

Planned future detectors include both ground-based detectors and a space-based detector.  Of the ground based detectors, LIGO-India~\cite{LIGO-India2013} is a third LIGO interferometer being constructed in India to increase the ability of the existing ground-based interferometer network to localize the sky location of a detected gravitational-wave source.  The current LIGO and Virgo detectors, called Advanced LIGO (aLIGO) and Advanced Virgo (adVirgo), will have another significant upgrade after which they will be called A+ and AdV+\cite{ASTROO2020}.  There are also next generation interferometers proposed including the Einstein Telescope~\cite{EinsteinTelescope2011} in Europe comprised of three 10-km long arms forming an equilateral triangle, and the Cosmic Explorer~\cite{CosmicExploer2019} in the United States with two 40-km long orthogonal arms.

The Laser Interferometer Space Antenna (LISA)~\cite{LISA2017} is a space-based detector planned to be launched by the European Space Agency.  LISA is a constellation of three satellites arranged in an equilateral triangle with a distance of 1 million km between each satellite. The constellation will orbit the Sun $20^\circ$ behind the Earth with each satellite on its own orbital path so that they also rotate once around the center of their configuration per orbit.  LISA will be sensitive to gravitational waves between the frequencies of $0.1\textendash100$ mHz. This frequency range lies between that of pulsar timing techniques and interferometric techniques. 

\emph{This review focuses on the state-of-the-art data analysis techniques currently used in the LIGO and Virgo interferometers.}  KAGRA began taking data in 2020 and will use many of the data analysis techniques reviewed here.

\section{Data}
\label{section:data}

LIGO and Virgo observe fluctuating laser power created by the changing relative length of their interferometer arms as a gravitational wave passes.  These power measurements are then calibrated into a strain timeseries sampled at 16,384 Hz in LIGO~\cite{Cahillane:2017vkb} and 20,000 Hz in Virgo~\cite{Acernese2018}.  These data are made publicly available through the Gravitational Wave Open Science Center (GWOSC)~\cite{GWOSC2019, GWOSCurl}. GWOSC curates full strain data for the events contained in the latest catalog release as well as strain data passing certain data quality cuts for full observing runs.  There are also many auxiliary channels recorded that monitor the behavior of the detectors and their environment, although GWOSC currently has limited auxiliary data available.  

Advanced LIGO and Advanced Virgo have completed 3 observing runs, labeled ``O" and a sequential number.  O1 took place between 12 Nov 2015 to 19 January 2016 and O2 took place between 30 November 2016 to 25 August 2017. O3 was broken into to parts: O3a took place between 1 April 2019 to 1 October 2019 and O3b took place between 1 November 2019 to 27 March 2019\cite{timeline}.  KAGRA joined LIGO and Virgo in taking science quality data during O3b.  O4 is expected to begin during the summer of 2022.  

A critical step to detecting gravitational waves is to understand the properties of the noise contained in its observation data.  Improperly modeled noise can lead to incorrect estimates of the significance of an event and systematic biases in parameter estimation.  The sensitivity of the strain data is limited at low frequencies ($<$ 10 Hz) by ground motion and at high frequencies ($>$ 100 Hz) by quantum (shot) noise.  Other noise sources include gravity gradients, mirror suspension thermal noise, Brownian noise from optics and their coatings, and thermo-optic noise from coatings~\cite{TheLIGOScientific:2014jea, Stuver2019}.  There are also high amplitude, narrow bandwidth features (spectral lines) caused by coupling to the AC power grid, mechanical resonances of the mirror suspensions, injected calibration lines, and noise from the detector control systems~\cite{PhysRevD.97.082002, O2speclines}.  The calibration lines and the power grid lines are removed from the data before detection algorithms are applied~\cite{davis2021ligo}.  The remaining lines must be carefully modeled especially for analyses that search for continuous sources which have a stable frequency of gravitational waves (see section \ref{sec:continuous}, and reference ~\refcite{Riles2017}).

The noise in the LIGO and Virgo detectors is approximately stationary.  However, transient artifacts called glitches, caused by environmental or instrumental disturbances, occur frequently.  These glitches are studied to understand how they impact the different searches for gravitational waves and to measure the statistical confidence of candidate detections.

Once data has been collected, the effects of glitches can be mitigated through exclusion of contaminated times from producing a candidate gravitational wave or by modeling and subtracting the glitch from the data.  The two commonly used exclusion methods are gating and vetoes.  Glitches can also be modeled using wavelets and MCMC (Markov Chain Monte Carlo) methods~\cite{Chatziioannou_2021} so that they can be separated from candidate gravitational-wave signals and allow for better source parameter estimation.  This method of glitch subtraction is effective but computationally expensive.

Gating is used to remove high-amplitude glitches from searches by multiplying the affected data by an inverse Tukey window to smoothly reduce the measured strain to zero at the beginning of the glitch and then back to the measured strain at the end.  Gates typically have a sub-second duration.~\cite{davis2021ligo}

Vetoes are times when the data are known to be contaminated and not to be used to produce candidate detections~\cite{TheLIGOScientific:2017lwt}.  Data is deemed to be contaminated and included as a veto when a glitch is observed both in the detector strain data and in an instrumental or environmental data stream (auxiliary channel) that is insensitive to an astrophysical gravitational wave.  These channels are known as “safe” channels and minimize the chance that what appears to be a glitch may be a true gravitational wave.  

There are categories of vetoes numbered 1 through 5, with category 1 being the most severe~\cite{TheLIGOScientific:2016zmo, davis2021ligo}.  Vetoes identified as category 1 include times where the detector is not in its nominal state/configuration, missing data or has poor calibration.  Category 2 contains times of well-understood contamination that are excluded from the data before being processed by individual searches.  Category 3 vetoes are statistically significant, but the coupling mechanism is less well understood compared to category 2 vetoes.  Category 1 and 2 vetoes are applied before triggers (times analysis methods identify as significant) are produced while Category 3 vetoes are applied after.  Only categories 1-3 are actively used for data analysis purposes, although their specific usage can vary.  Vetoes typically have a duration of a second to a few seconds.

While the data are recorded as a timeseries, there are features that become apparent when represented in the frequency domain.  For example, transforming the timeseries data using a Fast Fourier Transform (FFT) will show the relative amplitudes and phases of sinusoids over a range of fixed frequency widths (bins).  A time-frequency representation can be assembled by applying the Fourier transform to successive periods of data. The high level of noise at low frequencies due to ground motion is readily apparent as are the spectral lines.  Data analysis techniques often take advantage of the representation of data in the frequency domain.~\cite{davis2021ligo}  Common data conditioning that is easily, but not exclusively, applied in the frequency domain includes equalizing the amplitude of the noise across the frequency range (whitening) and excluding specific frequency ranges (band passing).

The wavelet transform is another frequency domain representation of timeseries data that uses an infinite orthonormal set of basis of functions, based on a ``mother function,'' that can scale to the duration of a feature.  This allows a time-frequency representation to be generated directly from the single application of the wavelet transform with broader time bins at low frequencies and narrower time bins at higher frequencies.  There are different mother functions that generate different wavelet families.  The wavelet family chosen for a particular application of the wavelet transform depends on the properties of the data.~\cite{davis2021ligo, WaveletReview}

\section{Compact Binary Coalescence} \label{sec:cbc}
The coalescence of compact binaries composed of neutron stars and stellar-mass black holes are currently the most promising sources of new gravitational-wave discoveries.
While there is a large population of compact objects below 1 M$_\odot$, the only objects compact enough to allow for detection by current ground-based detectors are black holes and neutron stars.
Since the first discovery of gravitational waves from a binary black hole (BBH) in 2015\cite{Abbott:2016blz}, 50 detections from compact binary coalescences have been made by LIGO and Virgo.
The 11 detections made during Advanced LIGO and Virgo's first and second observing run were reported in the catalog GWTC-1.\cite{LIGOScientific:2018mvr}
This included 10 signals from BBH coalescences as well as the first signal from a binary neutron star (BNS) coalescence, GW170817,\cite{TheLIGOScientific:2017qsa} which was also the first
joint detection of gravitational waves and electromagnetic radiation.
Recently, 39 additional detections were reported from the first part of the third observing run (O3a) in the catalog GWTC-2.\cite{Abbott:2020niy}
Among these were the second signal from a BNS (GW190425), two signals from BBHs with very asymmetric masses (GW190412 and GW190814), and a signal from the merger of two heavy black holes that formed an intermediate-mass black hole (GW190521).
With these detections, we are probing gravity in the strong-field regime, establishing population parameters of BBH systems, placing constraints on neutron-star matter as well as impacting many other fundamental questions in physics and astrophysics\cite{MLPA-Isi}.

\subsection{Potential Sources}

Compact binary systems evolve according to general relativity.
As the system completes orbits, energy and angular momentum are lost due to gravitational radiation.
This causes the orbit to shrink and the two objects to slowly inspiral.
The emitted gravitational waves increase in frequency and amplitude in a characteristic chirp-like pattern, until the two objects merge.
Depending on the masses of the objects, the nature of the merged object can either be a hypermassive neutron star, a supramassive neutron star, or a single perturbed black hole from prompt collapse~\cite{Baumgarte1999,Dietrich2020,Bernuzzi2020}. In the case of the single perturbed black hole, it will radiate gravitational waves as a superposition of quasinormal ringdown modes that will be exponentially damped~\cite{Regge1957,Teukolsky1973,Leaver1985,Kokkotas1999}. 
In the case of the hypermassive or supramassive neutron star, the black hole formation is delayed and gravitational waves with characteristic frequencies depending on the properties of the remnant are emitted~\cite{Bauswein2011,Takami2014,Bernuzzi2015,Tsang2019}.
It is even possible for for a massive neutron star remnant, rather than a final black hole, to form from configurations with small total masses.

The amplitude and phase evolution of the emitted gravitational waves will depend on intrinsic and extrinsic parameters of the binary.
In the simplest case, four intrinsic parameters are needed including the component masses $m_1$ and $m_2$, where we choose the convention $m_1 \ge m_2$ and the spin angular momenta of the two binary components $\vec{S}_{1,2}$.
Typically, we represent these as dimensionless spin vectors  $\vec{\chi}_{i} = c \vec{S}_{i}/\left( G m_i^2 \right)$.
In addition, seven extrinsic parameters are needed to describe the location and orientation of the binary with respect to the observer including luminosity distance $D_L$, right ascension $\alpha$, and declination $\delta$, the binary’s orbital inclination $\iota$ and polarization angle $\psi$ and time $t_c$ and phase $\phi_c$ of coalescence measured in the frame of the observer.
A binary is said to be “face-on” if $\iota=0$ or $\iota=\pi$ and “edge-on” if $\iota=\pi/2$.
The polarization angle $\psi$ is needed to fully specify the radiation frame and specifies the orientation of the gravitational wave polarization axes relative to the detector arms.

Other parameters may modify the gravitational wave signal but are not regularly included in searches for compact binaries. For example, two additional parameters are needed to describe the eccentricity of the system: the magnitude $e$ and the argument of periapsis. But we typically assume that the radiation reaction has sufficiently circularized the binary orbit and that eccentricity is not a large effect in the sensitive band of ground-based detectors.
Nevertheless, dynamically formed binaries may have residual eccentricity as the signal enters the sensitive band of the detector so these systems have been the target of unmodelled searches of previous observing runs, but with no candidate events reported~\cite{Salemi2019}. More information about searches for eccentric systems can be found in Sect.~\ref{section:other_longdur} and~\ref{section:burst_cbc}. As another example, the internal
structure of neutron stars effects how the stars can be deformed via tidal interactions. This is quantified by the dimensionless tidal deformabilities $\Lambda_{1,2}$ of each binary component. We typically assume the tidal term has a small affect on the waveform in the most sensitive frequency band and is not needed for detection. However, systems with extreme equations of states could be missed with current searches~\cite{Cullen:2017oaz, Hinderer:2009ca}.

The full gravitational waveform for compact binary coalescence is described by three distinct phases: inspiral, merger and ringdown. The low frequency inspiral signal, also called a chirp, is characterized by monotonically-increasing frequency and amplitude as the orbital motion radiates away energy and the orbit shrinks. The inspiral signal for a binary with non-spinning objects can be well-modeled with post-Newtonian theory. Then the two polarizations are given by
\begin{eqnarray}
h_+(t) = -\frac{1+\cos^2\iota}{2}\left( \frac{G\mathcal{M}}{c^2D} \right) \left( \frac{t_c - t}{5G\mathcal{M}/c^3} \right)^{-1/4}\cos\left[ 2\phi_c + 2\phi\left(t-t_c;M,\mu \right) \right] \\
h_\times(t) = -\cos\iota \left( \frac{G\mathcal{M}}{c^2D} \right) \left( \frac{t_c - t}{5G\mathcal{M}/c^3} \right)^{-1/4}\sin\left[ 2\phi_c + 2\phi\left(t-t_c;M,\mu \right) \right]
\end{eqnarray}
where $D$ is the distance to the source, $\mathcal{M} = \mu^{3/5}M^{2/5}$ is the chirp mass with $M=m_1+m_2$ the total mass and $\mu=m_1m_2/M$ the reduced mass. The coalescence time $t_c$ and phase $\phi_c$ are defined to set a reference time and phase, and by convention, mark when the inspiral waveform ends.
In the case of binary systems with spinning component objects, the expressions for $h_{+, \times}(t)$ will have an additional dependence on the three components of the spins and will no longer differ by an amplitude scaling that depends only on $\iota$ and a constant phase shift.
The short merger phase marks the point at which the two compact objects start to coalesce and the peak of gravitational wave emission. This process is highly nonlinear and is modeled with numerical relativity simulations.
The merger is followed by a  high frequency ringdown phase that occurs after the objects have merged to a single object, likely a black hole. From black hole perturbation theory and numerical simulations, the ringdown signal is modeled as a superposition of quasinormal modes that decay exponentially with time.

\subsubsection{Binary Neutron Stars}

With our searches for gravitational waves from binary neutron star systems, we allow for the broadest definition of possible neutron star component masses~\cite{ozel_2012, Lattimer:2012nd, Kiziltan_2013, Ozel:2016oaf}. Based on observation and standard astrophysical scenarios, remnant neutron stars can have a minimum mass $\sim$1 M$_\odot$~\cite{Suwa:2018uni}.
From a theoretical perspective, requiring that the equation of state satisfies causality and our knowledge of nuclear matter at low densities provides a loose upper limit of $\lesssim 3\,\,\mathrm{M}_\odot$~\cite{Kalogera_1996, PhysRevLett.32.324, Linares_2018}. Regarding the range of angular momentum~\cite{O_Shaughnessy_2005, Hessels:2006ze}, we consider the distributions observed for pulsars in binaries with $\left|\chi_{1,2}\right|\lesssim0.04$ and add a bit more to allow for additional uncertainty~\cite{Kramer_2009}. Thus searches for gravitational waves from binary neutron stars to-date~\cite{Abbott:2020niy} have covered $m_{1,2}\in \left[1,3\right] \mathrm{M}_\odot$ with $\left|\chi_{1,2}\right| \le 0.05$.

\subsubsection{Binary Black Holes}

Stellar evolution models predict that black holes may exist with a minimum mass down to 2 M$_\odot$ and a maximum mass up to 100 M$_\odot$ or potentially higher. However, primordial processes could lead to the formation of binary black holes over a much wider mass range, including subsolar mass black holes. Regarding the range of angular momentum~\cite{fabian_2012, Gou:2011nq, McClintock_2011}, the relativistic Kerr bound allows $\left|\chi_{1,2}\right| \leq 1$~\cite{misner1973}. Thus standard searches for gravitational waves from binary black holes to-date~\cite{Abbott:2020niy} have covered $m_{1,2}$ between $3\,\,\mathrm{M}_\odot$ and $\sim100-400\,\,\mathrm{M}_\odot$ with $\left|\chi_{1,2}\right| \le 0.999$, as close to the Kerr limit as allowed with current waveform approximants.

Sub-solar mass primordial black holes could make up a fraction of dark matter.
Searches for these sub-solar binaries with at least one component between 0.2 and 1.0 M$_\odot$ have been performed with no detections to-date~\cite{Authors:2019qbw}.

Black holes with masses between $10^2$ and $10^5$~M$_\odot$ are classified as intermediate-mass black holes~\cite{Miller:2003sc}, and bridge the gap between stellar black holes and supermassive black holes. They may provide the missing link to explain the formation of supermassive black holes~\cite{Ebisuzaki:2001qm, Mezcua2017, Koliopanos2017}. The recent gravitational-wave detection GW190521~\cite{Abbott:2020tfl} is consistent with a binary black hole merger that has a final remnant mass of $142^{+28}_{-16}$ M$_\odot$, classifying it as an intermediate-mass black hole.

In the sensitivity band of Advanced LIGO and Virgo, signals from intermediate-mass black hole mergers will be short duration (tenths of seconds). There are very few cycles observable as the systems will merge at frequencies of 10s of Hz, so detection of these signals can be made by modeled matched-filter searches and also by weakly-modeled transient burst searches (see Sect.~\ref{section:burst_cbc}). A recent search for gravitational waves from intermediate mass black holes placed upper limits on the merger rate density for sources with total masses $M\in\left[120,800\right]\,M_\odot$~\cite{Aasi:2014iwa}.

\subsubsection{Neutron-star Black-hole Binaries}
Neutron-star-black-hole systems are thought to be efficiently formed either through the stellar evolution of field binaries or through dynamical capture of an neutron star by a black hole. Though no neutron-star-black-hole systems are known to exist, likely progenitors have been observed, including Cyg X-3~\cite{Belczynski:2012jc}.

To-date, standard searches for gravitational waves from neutron-star-black-hole systems~\cite{Abbott:2020niy} have covered neutron star masses $m_{2}\in\left[1,3\right]\,\mathrm{M}_\odot$ and black hole masses $m_{1}$ are typically chosen so that the mass ratio does not exceed $q=m_2/m_1=0.01$.

\subsection{Analyses}
Searches for modeled sources of gravitational radiation typically use matched-filter-based analyses.
Matched-filtering is necessary for the detection of compact binary signals for which the signal energy is spread over a long time interval. For the analyses presented here, this is typically $\lesssim$ 1 minute in the sensitivity band of Advanced LIGO and Virgo. The method correlates a waveform model with data over the detectors' sensitive bands to extract signals from detector noise. The method is optimal when a known signal waveform is in stationary Gaussian noise.

The matched-filter output of a data stream $s(t)$ with a filter template $h(t)$ is given by the noise-weighted cross-correlation
\begin{equation}
    \langle s, h \rangle= 4 \Re\int^\infty_0 \frac{\tilde{s}(f) \tilde{h}^*(f)}{S_n(f)} df
\end{equation}
where $S_n(f)$ is the one-sided power spectral density and $\Re$ denotes the real part. The data stream $s(t)$ may contain just noise or noise and a signal. To construct the signal-to-noise ratio, the norm of each template is computed, using systems at a distance of 1 Mpc, as $\sigma^2 = \langle h,h \rangle$. Then the matched filter signal-to-noise ratio is defined as
\begin{equation}
    \mathrm{SNR} = \frac{\langle s, h \rangle}{\sigma}.
\end{equation}

Template banks of waveforms are constructed via geometric~\cite{Babak2006,Cokelaer2007}, stochastic~\cite{Harry2009,Privitera2014}, or hybrid methods~\cite{Roy2017,Roy2019}. Traditionally, searches have been performed for (anti)-aligned spin systems and hence template placement is in $\{m_1, m_2, \chi_1, \chi_2 \}$-space. Typically, the placement is such that the loss in matched-filter SNR caused by the bank's discrete nature is $\lesssim 3\%$. However, sub-threshold and focused-searches often choose $\lesssim 1\%$.

Template waveforms for matched filtering use precise waveform models of compact binary coalescence~\cite{Taracchini2014, Pan2014, Hannam2014, Khan2016}.
Waveforms have been developed combining various techniques to model the two-body dynamics and gravitational wave emission.
For lower mass binaries (component masses $\lesssim$ 3 M$_\odot$) post-Newtonian theory is suitable for describing the inspiral waveform.
For higher mass binaries (component masses $\gtrsim$ 3 M$_\odot$~\cite{Ajith:2007xh}), the merger and ringdown phases will occur at lower frequencies, potentially within the detectors' sensitive band.
Thus, templates also need to include these phases.
For this, the effective-one-body formalism~\cite{Damour:2008yg} and numerical relativity~\cite{Baumgarte:2002jm, Duez:2018jaf} are needed to describe the full waveform.
The first relativistic corrections to the Newtonian dynamics were obtained almost a century ago~\cite{Lorentz1917, Einstein1938}.
More recently, waveforms have been provided through various techniques including
higher-order post-Newtonian calculations~\cite{Blanchet1995,Blanchet2004,Blanchet2014}, analytical techniques to model the relativistic two-body dynamics and gravitational waves~\cite{Buonanno2000,Buonanno1999, Damour2008, Damour2009, Barausse2010}, and numerical-relativity
simulations~\cite{Pretorius2005,Campanelli2006,Baker2007,Hinder2013,Mroue2013,Husa2016}.
Waveform models that accurately describe the two-body dynamics and the emitted gravitational waves are used to construct the set of matched filters~\cite{Taracchini2014,Purrer2016}.

Each matched-filter search identifies “triggers” from individual interferometer data streams.
These triggers mark GPS times where the SNR exceeds a preset SNR threshold (typically SNR$>4$).
As described in Sect.~\ref{section:data}, detector data often contain glitches that can falsely produce high SNR.
Thus, periods of bad data quality are often removed from the analyses using vetoed times or triggers from these times are penalized in the final candidate ranking list.
After the initial trigger, searches employ a pipeline of statistical tests for further candidate vetting.
Search pipelines typically maximize the SNR of each template over time for short windows ($\sim 1$\,s) and only record
triggers passing the threshold for every data stream.
Various types of clustering algorithms are used to reduce trigger lists.
Often a coincidence test is applied requiring that a trigger from the same template is found
within the inter-site light travel time, plus a small window for uncertainty in the arrival time of weak signals.
Recently, single-detector triggers have been analyzed, relaxing the requirement for multi-detector coincidence.

Search pipelines then employ a set of consistency tests that may include information about the time delay and phase differences between candidates in each detector, the sensitivity of each detector, the time-frequency morphology of the candidates and expectations for different astrophysical population models.
The parameters derived from each of these tests are then used to construct ranking statistics for each of the candidates.
Below we provide more details on each of the matched-filter-based search pipelines currently used for searches for gravitational waves from
compact binary coalescence in LIGO and Virgo.

Search pipelines may run in either low-latency mode on streaming gravitational-wave data or offline mode on archival gravitational-wave data.
The goal of low-latency analyses is to provide rapid alerts of candidate gravitational-wave signals, particularly for BNS signals which may be accompanied by an EM counterpart.
GCN Notices from low-latency gravitational-wave pipelines are currently released with a typical latency of a few minutes~\cite{Magee_2021}.
Recently, several pipelines have also demonstrated the ability to make detections of the pre-merger inspiral of long-lived BNSs, potentially enabling more effective EM followup~\cite{Sachdev_2020, Nitz:2020vym}.
A mock data challenge was recently completed with two pipelines demonstrating the ability to produce pre-merger (early warning) detections of binary neutron star signals and send alerts to partner facilities before merger~\cite{Magee_2021}.

\subsubsection{GstLAL}
The GstLAL-based inspiral pipeline~\cite{Messick2019,Sachdev2019}, built upon the GstLAL library~\cite{Cannon2021, GstLAL}, utilizes the GStreamer framework~\cite{GStreamer} and the LIGO Algorithm Library~\cite{LALSuite}. It runs in both low-latency and offline modes.

This pipeline performs matched-filtering in the time-domain and employs the LLOID method~\cite{Cannon2012} which
combines singular value decomposition (SVD)~\cite{Cannon2010,Cannon2011}
with multi-banding to construct a reduced set
of both matched filters and samples. Typically, candidates with $\mathrm{SNR}>4$ are kept for further vetting.

The pipeline utilizes a multidimensional likelihood-ratio ranking statistic $\mathcal{L}$ to identify gravitational-wave candidates~\cite{2015arXiv150404632C, Messick2019, Sachdev2019, Hanna2020, Abbott:2020niy}. The statistic incorporates information including the SNRs of the triggers, a time-domain signal-consistency test~\cite{Messick2019}, timing and phase differences between coincident triggers in each detector~\cite{Sachdev2019, Hanna2020}, the time-averaged volumetric sensitivity of each detector, the signal population model~\cite{Cannon:2012zt} and the probability that a signal trigger is recovered by a waveform template given its SNR~\cite{fong_2018}, and the background noise model described below. Additionally, single-detector candidates when only one detector is operating are ranked with additional data quality information from machine-learning based predictions (iDQ)~\cite{Essick:2020qpo, Godwin:2020weu}.

The pipeline computes the background of the search by first splitting the template bank into many bins with similar time-frequency evolution. Statistics from non-coincident triggers when more than one detector are operating are histogrammed in each bin. This set of triggers characterize the noise background of the search, in each bin. Final rankings are determined after marginalizing over all the bins. Monte Carlo samples from these background $\mathcal{L}$ distributions are drawn to determine the mapping between $\mathcal{L}$ and the false-alarm probabilities of the candidates~\cite{Cannon:2012zt}.

The signal model used in determining $\mathcal{L}$ for each candidate is determined by assuming a uniform-in-volume distribution of sources and a maximum of 10\% loss in SNR due to waveform mismatch~\cite{Cannon:2012zt}.

\subsubsection{PyCBC}
The PyCBC-based inspiral pipeline~\cite{Usman2016,Nitz2017,DalCanton2014} performs matched-filtering in the frequency domain. It runs in both low-latency and offline modes.

Data during times flagged by category 2 veto flags are removed from PyCBC searches. The pipeline requires triggers in more than two detectors using a multi-detector coincidence test~\cite{Davies2020}. These candidate events are then ranked using a multi-detector ranking statistic~\cite{Davies2020} that incorporates information including the SNRs of the triggers,
signal consistency tests such as the time-frequency $\chi^2$ veto~\cite{Allen2005} and a test to identify excess power at high frequencies~\cite{Nitz:2017lco}, the network sensitivity for each detector,
phase-time-amplitude consistency checks and a comparison of the properties of the event against expected signal and noise populations. The signal model is incorporated as signal prior distributions that model the amplitudes, phases, arrival times in detectors with differing sensitivities. 

The pipeline computes the background of the search using the method of time-shifted single-detector triggers. These are performed using intervals of 0.1\,s using a many-detector shifting procedure described in Ref.~\refcite{Davies2020}. Triggers from apparent gravitational-wave signals are iteratively removed from the time-shifted analyses and only times from when all detectors were observing are included to avoid biasing the significance estimation.
For each candidate with a measured ranking statistic, a false alarm rate is computed by comparing the rate of coincident noise events from the time-shift analyses that had a higher ranking statistic.

Recently, the results from a focused search for binary black hole coalescences~\cite{Nitz2020} was presented~\cite{Abbott:2020niy}.
This search uses the detection statistic presented in~\cite{Nitz2019,Davies2020} which includes a number of tuning choices to reject triggers that do not well match the
filter waveforms, and also includes a template weighting
implementing a prior that signals detectable in any given
range of SNR are uniformly distributed in chirp mass.

\subsubsection{MBTA}

The multi-band template analysis (MBTA) pipeline~\cite{Adams2016} runs primarily as a low-latency coincident analysis pipeline for the detection of gravitational waves from compact binary coalescences.

The MBTA pipeline identifies triggers from individual interferometer data streams and then performs a coincident analysis using time and exact template match coincidence.
To reduce the computational cost of the matched filtering, MBTA splits the matched filter
across two (or more) frequency bands. The boundary frequency between the low
frequency (LF) and high frequency (HF) bands, $f_c$ is selected so that the SNR is shared roughly equally between the low and high frequency bands,
typically $f_c\approx100$~Hz, for advanced detectors.
On average, this procedure loses no SNR compared to a matched filter performed with a single band
analysis .

Due to the use of multibanding, MBTA employs a computationally inexpensive signal consistency test, the $\chi^2$ cut~\cite{Allen2005}.
Additionally, a matched-filter timeseries signal consistency test is also employed, to supplement the $\chi^2$ cut. This test exploits the fact that the matched filter timeseries for a gravitational wave signal will have a single narrow peak, while noise will have a timeseries with a broader peak and multiple maximums around the central feature~\cite{Guidi2004,Shawhan2004}.

In low-latency, the pipeline computes the background of the search by making every possible coincidence from single detector triggers in recently collected data, and computes the probability of a pair of triggers passing the coincidence test.
The significance of each candidate is estimated by calculating the false alarm rate as the expected rate of coincidence triggers from background triggers that have an equal
or larger combined SNR than the candidate.

\subsubsection{SPIIR}
The summed parallel infinite impulse response (SPIIR) pipeline~\cite{Luan2012, Hooper2012, chu_2017} runs primarily as an online low-latency coincident analysis on time-domain data.
SPIIR applies time-domain summed parallel infinite impulse response (IIR) filters to approximate matched-filtering results with high accuracy and, in theory, zero latency.
For online low-latency analyses, this method is more computationally efficient than traditional Fourier methods, achieving a median online latency during O3 of $\sim 9s$~\cite{Chu2020}.
Additionally, parallelization of this algorithm is straightforward using Graphics Processing Units (GPUs) to accelerate both filtering~\cite{Liu2012, Guo2018} and coherent candidate selection~\cite{Luan2012}.

GPU acceleration also enables SPIIR to apply an online coherent search that coherently adds SNR output for each sky direction from the detector network to form a detection statistic~\cite{chu_2017, Liu2012}.
This detection statistic, called the coherent network SNR, is based on the maximum likelihood ratio principle~\cite{Guo2018}.
To address the computational challenge of finding triggers in real-time, high SNR candidates from each detector are pre-selected and then the coherent responses from other detectors are found for each possible sky direction.
The pipeline ranks the triggers by the coherent network SNR and the average $\chi^2$ values from individual detectors.
It computes the background of the search by performing 100 time-shifts over two weeks.
The K-nearest-neighbor (KNN) technique is used to estimate the significance for triggers~\cite{Chu2020}.

\section{Bursts} \label{sec:bursts}

Many astrophysical processes are expected to emit transient signals including merging compact binary systems consisting of black holes and/or neutron stars, core collapse supernovae, magnetars, and cosmic string cusps. The key strategy of  transient unmodelled burst searches is to search for gravitational-wave signals without making assumptions on the morphologies of these signals. This approach allows the burst searches to be sensitive to a wide range of signal morphologies and astrophysical models for which robust models are not available.
The burst searches include analyses that do not make assumptions either on the source sky location or the gravitational wave arrival time, labelled as “all-sky searches”, and targeted searches that involve information from multimessenger observations such as supernovae or gamma-ray burst (GRB) searches.
Burst searches can also detect and reconstruct well modelled transient signals (e.g. due to compact binary mergers) with the flexibility to allow for the existence of unexpected characteristics of the signal. In fact, parametrized models of the detected emissions from compact binary coalescences may not always accurately cover all the gravitational wave emission features such as the ones in the case of orbital eccentricity, misaligned spins and post-merger emission from neutron star remnants.

\subsection{Unmodelled pipelines} 

The different algorithms to search for unmodelled transient gravitational waves are based on the identification of statistically significant excess power in the time-frequency (TF) representation of the whitened data. The (frequency-domain) noise-scaled whitened data is defined as $\mathbf{w}[i] =\mathbf{x}[i]/\sqrt{S_k[i]}$, where $S_k[i]$ is the power spectral  density, and $\mathbf{x}[i]={x_1[i],...,x_k[i]}$ the timeseries, where $k$ refers to the detector and $i$ to the data sampling.  
The timeseries observations made by the interferometer, $\mathbf{x}[i]$, are defined as:
\begin{equation}
\mathbf{x}[i]= \mathcal{F}\mathbf{h}[i] +\mathbf{n}[i] 
\end{equation}
where $\mathbf{n}[i]$ is the detector noise realization, $\mathbf{h}[i]=\big[h_{x}[i],h_{x}[i]\big]$ is the gravitational wave signal composed of its two polarizations, $h_{+}$ and $h_{\times}$, and $\mathcal{F}$ is the network  antenna pattern matrix:

\begin{equation}
  F=
\begin{bmatrix}
F_{1+}(\theta,\phi) & F_{1\times}(\theta,\phi)\\
 \cdots & \cdots \\
F_{k+}(\theta,\phi) &F_{k\times}(\theta,\phi)\\
\end{bmatrix}
\end{equation}
where $\theta$ and  $\phi $ are the sky position coordinates.  

A common  approach \cite{Tinto, Chatterji_2006, Klimenko_2005} to the detection and reconstruction of gravitational-wave signals, defines the two hypotheses: $H_0$  “data when a gravitational-wave signal is \emph{absent}” and $H_1$  “data when a gravitational-wave signal is \emph{present}." Assuming the noise as stationary Gaussian white noise not correlated between detectors, the corresponding probability densities can be determined as:
\begin{equation}
p(\mathbf{w} \mid H_0) = \prod_{i=1}^{N} \frac{1}{\sqrt{2 \pi} \sigma} exp \left( - \frac{\mathbf{w}[i]^2}{2 \sigma^2} \right) 
\end{equation}

\begin{equation}
p(\mathbf{w} \mid H_1) = \prod_{i=1}^{N} \frac{1}{\sqrt{2 \pi} \sigma} exp \left( - \frac{(\mathbf{w}[i] -\boldsymbol{\xi}[i])^2}{2 \sigma^2} \right) 
\end{equation}
where $\boldsymbol{\xi}[i]=F[i]\mathbf{h}[i]$ is the detector response to a gravitational-wave signal.

The likelihood ratio is then: 
\begin{equation}
\Lambda (\mathbf{w}, \Omega) = \frac{p(\mathbf{w} \mid H_1(\Omega))}{p(\mathbf{w} \mid H_0(\Omega)} 
\end{equation}
where  $\Omega$ are the parameters of the sky position for the signal source. The maximization of the likelihood ratio over the possible sky position provide the solution for the gravitational-wave waveforms, $h_{+}$ and $h_{\times}$.

Multi -detector analysis can be performed using coincident or coherent approach. Coincident methods require events from individual detectors to be identified within a fixed time window (with duration compatible with the light travel time between detectors) and with similar morphology. On the other hand, coherent methods combine data streams from multiple detectors and builds a ranking statistic as a coherent sum over the detector responses.

To estimate the statistical significance of a foreground gravitational-wave candidate, the standard procedure calculate its false alarm probability by comparing its ranking statistic to the distribution of the ranking statistic for the expected background of accidental trigger. This is obtained by repeating the analysis on many instances of network data where non-physical time-shifting has been introduced; this breaks coherence in the network data and produces triggers that are due only to random coincidences in detector noise without any contribution from real signals. The expected background distribution generated with the time-shift methodology is able to take in to account non-Gaussian and non-stationary features of the  considered data timeseries. This procedure is highly computational demanding, involving the analysis of many years of equivalent data.
  
\subsubsection{Coherent WaveBurst (cWB)}
The Coherent WaveBurst (cWB) algorithm \cite{Klimenko2016} is based on a coherent maximum likelihood approach applied to the multi-resolution time-frequency representation of the whitened timeseries of the detectors’ data through the fast discrete Wilson-Daubechies-Meyer (WDM)  transformation \cite{WDM_Necula}. Using a linear combination of wavelet bases at different resolutions  allows for a more complete representation of the signal.
The triggers are then identified by clustering the pixels that pass the threshold on the excess power over all the network interferometers.  For the selected cluster of pixels, the likelihood statistic is built and the maximization is determined with a loop over all possible sky positions. 
cWB’s peculiarity is the possibility to promote a selection of clusters with a given pattern. For example, patterns in which frequency increases with time (a chirping structure) is especially suitable for most of the CBC sources. 
The cWB event ranking statistic is proportional to the coherent signal-to-noise ratio (SNR) across the network of detectors. cWB then estimates the network correlation, defined as the ratio between the coherent energy to the total energy (sum of coherent and incoherent energy or residual noise). The network correlation is used to better discriminate gravitational-wave signals as it is expected to be close to  unity for genuine gravitational-wave signal and $<<1$ for non stationary noise fluctuations (glitches).

\subsubsection{BayesWave}
The BayesWave \cite{Cornish_BW_2021, Cornish_BW_2015} algorithm considers the data from an interferometer $k$ ($x_{k}$) as the linear combination of interferometer response to gravitational wave signals ($h_{k}$), Gaussian noise ($n_{k}$), and transient glitches ($g_{k}$): $x_k = h_k+n_k+g_k$. To model the properties of a gravitational wave $h$ from data $x$ under model $M$ using Bayesian statistics, it calculates the posterior distribution function $p(h\mid x,M)$:

\begin{equation}
p (h  \mid x,M )=  \frac{p(x \mid  h,M) p(h\mid M)}{p(x\mid M)}
\end{equation}
where $p(h \mid M)$ is the prior likelihood for $h$ in model $M$, $p(x \mid h, M)$ is the likelihood of the data for the given waveform, and $p(x\mid M)=\int p(h \mid M)p(x \mid h, M)$ is the marginal likelihood,  or  evidence,  for  the  model $M$ under  consideration.

The  signal and glitch models are built as the linear  combination  of  sine-Gaussian Morlet-Gabor wavelets, with a  number of  wavelets independently optimized for the two models by a reversible jump Markov Chain Monte Carlo. 
The algorithm considers three possible  hypotheses: the data contain only Gaussian noise,  the data contain Gaussian noise and glitches, or the data
contain Gaussian noise and  gravitational wave signal (which requires at least one wavelet to be coherently projected onto the network).
The Bayes factor between the latter two models is used as detection statistic.
This algorithm  produces posterior distributions for the model parameters as the  reconstructed waveform signal and source sky position.

\subsubsection{Omicron-LIB (oLIB)}
The Omicron-LIB (oLIB) \cite{oLIB_Lynch}  algorithm is an hierarchical pipeline that consists of a initial incoherent stage  in which Omicron \cite{Robinet_Omicron} identifies excess power  in a single interferometer’s data stream  through the Q-transform \cite{QTransform, Chatterji_QTrasform}.  Then only triggers with similar central frequency and quality factor  across detectors and with temporal difference consistent with  time-of-flight  window, are selected.

The selected events are passed onto LALInferenceBurst (LIB) algorithm based on LALInference, that models signals with sine-Gaussian templates, depending on parameters such as frequency,  quality factor, amplitude, and sky position. This stage of the algorithm computes two Bayes factors (natural logarithm of the ratios of the evidence of two hypotheses): the first as a gravitational-wave signal  vs. a Gaussian noise signal, and the second as a coherent gravitational-wave signal vs. an incoherent noise transient. The joint likelihood ratio of these two  Bayes  factors, defines the ranking  statistic of the search to estimate the gravitational-wave detection significance.

\subsection{Short duration all-sky searches} 

As discussed previously, gravitational-wave bursts can be generated by a wide variety of astrophysical sources, such as merging compact binary systems, core-collapse supernovae of massive stars, coalescence of binary neutron stars, pulsar glitches, and cosmic string cusps. These kinds of signals are expected to generate signal with frequency in the range from several Hz to a few kHz, and duration in range from $10^{-3}$ s to tens of seconds.

cWB, oLIB and BayesWave have been employed to perform gravitational-wave searches in this parameter space, for the past LIGO Virgo data takings. The use of multiple search algorithms with different performances for specific classes of gravitational-wave signals can improve coverage of the wide  parameter space, increase overall detection efficiency, and provide independent validation of the results. 

The  cWB search is divided in two different frequency ranges which cover the low (32 - 1024 Hz) and high frequency (up to 4kHz) bands.
Burst searches' sensitivity can be strongly affected by the fact that data can been polluted by glitches due to instrumental  and  environmental  noise  artifacts. To minimize their impact, in the LIGO and Virgo analysis, cWB triggers have been divided into bins, with different signal waveform  features. For O1 and O2 analyses, the triggers in the low frequency band search  have been divided in two classes. In particular one of these classes has included signals with most of their energy in a frequency bandwidth of few Hz, and with a small quality factor, which separated triggers caused by non-stationary power spectrum lines and `blip'-glitches\cite{Cabero_2019,Davis_2020} (a  type  of  low  frequency, short duration terrestrial glitch which cannot be effectively vetoed) from triggers that may be astrophysical in origin. For the analysis of O1 and O2 data collections BayesWave pipeline is employed as a follow-up of the cWB search by analysing cWB triggers in the low frequency bands exceeding a fixed detection statistic.

oLIB search covers the low ($ 32 - 1024$ Hz)  and  high ($  1024  -  2048$ Hz) band  frequency range; then triggers identified candidate events are classified in bins depending on  the  quality factor of the signal.

Unmodelled searches for short-duration gravitational-wave transients in  O1 and O2 data reported null results, apart  from  the  known  BBH  signals detected by CBC template searches; upper  limits  on  the  $90\%$  confidence  intervals  for the  gravitational-wave rate-density have been estimated through a full simulation campaign, and they are reported in Ref. \refcite{Abbott:2019prv}.

\subsubsection{Search: CBC related searches for Binary Black Holes, Intermediate Mass Black Holes (BBH-IMBH), and eccentric binary systems} 
\label{section:burst_cbc}

Burst unmodelled transient algorithms can also complement the template searches for CBC sources; these algorithms can be sensitive to signals with features due to  higher-order modes, high mass ratios, misaligned spins,  eccentric orbits  and,  possible  deviations  from  general  relativity.
Moreover, as discussed in section \ref{sec:cbc}, IMBH binary systems are expected to merge at low frequencies resulting in few signal cycles being in LIGO's and Virgo's sensitive bandwidth. We leverage the strength of burst analyses' lack of waveform assumptions by applying them to these well-modelled sources to compliment the CBC search.

These searches contributed to the detection of the first gravitational waves signal GW150914; it was in fact detected and reported within three minutes of data acquisition by low-latency unmodelled searches  \cite{Abbott:2016blz}. In Ref. \refcite{TheLIGOScientific:2016uux} the analysis of GW150914 with minimal assumptions is discussed, reporting the significance estimation of the detection, waveform reconstruction, and consistency with the theoretical signal from a binary black-hole merger.

Unmodelled searches for CBC signals targeting low-mass and high-mass BBH systems have been performed for O3 Advanced LIGO-Advanced Virgo data using two different configurations of the cWB analysis \cite{Abbott:2020niy, Abbott:2020tfl, IMBBH_cWB_Szc}. 
Minimally-modeled low-mass searches require the frequency of a trigger to be increasing in time (chirping time-frequency pattern). Moreover,  unique data quality vetoes are defined for this search to limit the consequence of non-stationary noise in the detectors’ bandwidth.

The cWB-BBH search is used also used to search for black hole  mergers  that  inspiral  in  eccentric  orbits, which are not fully modeled by theoretical predictions. Simulation studies have showed \cite{Salemi:2019owp} that the detection capability is independent of the eccentricity at the time the binary enters Advanced LIGO's and Advanced Virgo’s frequency band at $\sim 10$ Hz.

Unmodelled algorithms are also employed to test general relativity and the consistency of the waveform reconstruction to templates models.
A transient signal strongly deviating from general relativity and missed by template searches could be found by unmodelled or minimally-modelled searches, as well as the specific features of the signals that are missed by templated models \cite{Abbott:2020niy}. 
Consistency tests of unmodelled waveform reconstruction ($h_{1}$) and maximum likelihood template-based waveforms ($h_{2}$) are based on the evaluation of the agreement of the two waveforms by the overlap of the two signals, defined as:
  \begin{equation}
\textbf{O} \langle h_1 , h_2 \rangle  = \frac{\langle h_1 \mid h_2 \rangle}{\sqrt{\langle h_1 \mid h_1 \rangle \langle h_2 \mid h_2 \rangle}}
   \end{equation}

The expected distribution of these match values is obtained by performing analyses on off-source data (data surrounding the event) adding waveforms from  the  template-based analysis.  The p-value of the detected events is given by the fraction of off-source match values that are below the on-source match value\cite{Abbott:2020niy}.

To test for possible deviation from general relativity, the algorithm analyzes the residual power in the data once  the best-fit template is subtracted
\cite{Abbott:2016blz,LIGOScientific:2019fpa,Abbott:2020jks}. Specifically, for each event, the BayesWave pipeline produces a distribution of possible residual coherent signals consistent with the data, that allows to estimate the 90\%-credible upper limit.  

\subsection{Long duration all-sky searches}
\label{section:other_longdur}
Different astrophysical phenomena are foreseen to emit transient gravitational-wave signals with intermediate time duration between a tens to a few hundreds of seconds and in a  wide range of expected frequencies. Among different astrophysical processes, promising sources are: fallback accretion (fallback of ejected mass in newborn neutron stars induces deformation and emission of gravitational waves until it collapses into a black hole), accretion disk instabilities (stellar material spiraling into a black hole can be origin of gravitational-wave emission), and nonaxisymmetric deformations in magnetars. 
Theoretical signal  predictions for those astrophysical processes, when available, cover a wide range of morphologies and are in some cases poorly known. 

Four unmodelled pipelines have been employed to search for long duration transient gravitational-wave signals in LIGO and Virgo data: the long-duration configuration of cWB, two different versions of the Stochastic Transient Analysis Multi-detector Pipeline (STAMP-AS) pipeline, and the X-pipeline Spherical Radiometer (X-SphRad) \cite{XSphrad}.

The STAMP-AS \cite{Thrane_STAMPAS} pipeline performs cross correlation of data from two interferometers to calculate coherent time-frequency maps of cross-power SNR with pixel size of $1s \times 1 Hz$. 
Pixels whose SNR exceed a certain threshold are selected to form triggers by two possible algorithms. The first one, Zebragard \cite{Prestegard_thesis, Abbott:2015vir}  is a seed-based clustering algorithm, that groups pixels that are within a certain distance from each others. The ranking statistic is defined as the quadratic sum of the pixels' SNR.
The other algorithm, Lonetrack \cite{Thrane_LoneTrack_2015} employs a hierarchical approach. It uses a seedless cluster algorithm on the selected pixels of single-detector spectrograms; it calculates an incoherent statistic integrating the signal power along many parametrized Bezier curves. Single-detector triggers whose incoherent statistic passes a threshold are then used to build coherent detection statistic through cross-power SNR.  In both algorithms, a loop over the possible sky positions is performed.

The cWB pipeline has a specific configuration for the long-duration gravitational-wave transient search, the relevant differences with respect to the short duration transient search configuration are the use of specific time-frequency map time resolutions with longer pixels in time and specific post trigger production cut on time duration of the triggers. 

X-SphRad pipeline \cite{XSphrad} is based on a cross-correlation algorithm in the spherical harmonic domain.  The cluster algorithm groups next-nearest-neighbor pixels on a time-frequency representation of the data; then the trigger events are ranked by the ratio of the power in the homogeneous polynomials of degree $l >0$ modes to that in the $l= 0$ mode.
The background estimation is performed for all pipelines employing the usual time-slides methodology.

Searches for long duration gravitational-wave signals were performed for first and second LIGO-Virgo observation runs in the frequency range $24-2000 Hz$; no significant events were detected (apart from the binary neutron star merger GW170817). The best sensitivity limits on the root-sum-square strain amplitude $hrss$ at $50\%$ detection efficiency estimated for those searches through simulation campaigns are 
$hrss_{ 50\% }= 2.7 \times 10^{-22} Hz^{-1/2}$
for a millisecond magnetar model and $hrss_{50\%} =9.6 \time 10^{-22} Hz^{-1/2} $ for eccentric compact binary coalescence signals. \cite{Abbott:2017muc, Abbott:2019heg}

\section{Multimessenger searches} \label{sec:mma}
Many astrophysical  sources of gravitational-wave transients, such  as coalescing binaries, supernovae and  gamma-ray bursts (GRBs), are expected to  be source of  multimessenger signals, emitting also electromagnetic (EM) signals or neutrinos.  
Then the joint analysis of signals due to different messengers can provide a more complete  knowledge  of  the  astrophysical  sources.

Multimessenger astrophysics searches can follow different approaches. Triggered or directed transient gravitational-wave searches plan to use information provided by other messengers of a common astrophysical source such as sky positions, timing, and source parameters to improve the search sensitivity and candidate event significance. Likewise gravitational-wave detections can trigger electromagnetic searches, in this case source sky position of gravitational-wave candidates are used  to  point  EM  observatories  looking for possible counterparts in EM spectrum. This follow-up campaign is possible thanks to  the online identification of gravitational-wave candidates and the distribution of  alerts by the LIGO and Virgo collaborations  \cite{LIGOScientific:2019gag} in low-latency mode.

Targeted searches have been set to look for gravitational-wave transients associated with short and long GRBs, using both modelled and unmodelled search methods; in these searches gravitational-wave data are analysed in the time around a GRB detection, requiring the consistency of gravitational wave and GRB source sky position; in that way a better sensitivity with respect to the all-sky gravitational-wave searches is achieved.
The unmodelled approach employs  the X-Pipeline \cite{Xpipeline_Sutton_2010, TargetGRB_Was} algorithm that looks  for  excess  power coherent across the network and requires  consistency with the sky localization and time window for each GRB. The PyGRB pipeline  \cite{PyGRB_Harry, PyGRB_Williamson} is instead based on a coherent matched filtering search; it  looks for  a  gravitational-wave  signal  due to the inspiral of a BNS or neutron start-black hole binary in the time window associated with an observed short GRB.
The results for targeted gravitational-wave searches associated with GRBs for first and second LIGO/Virgo data collections \cite{Abbott:2016cjt, Authors:2019fue}, has involved the analysis of the 41
GRBs in the O1 data,  98 GRBs using the unmodelled method, and 42 using the modeled  method in the O2 data; these searches detected the already discussed GW170817A. The analysis of O3a data \cite{Abbott:2020yvp} has covered 105 GRBs for unmodelled method and  32 GRBs for modeled methods that targets BNS, finding no significant evidence for gravitational-wave signals associated with the GRBs.

Triggered or directed transient gravitational-wave searches plan to use information provided by common astrophysical source as sky positions and timing, to improve search sensitivity and candidate significance. Likewise, gravitational-wave detections can be used to  point  EM  observatories  and  search  for  the  EM signatures \cite{LIGOScientific:2019gag}.

In 2017, the first combined detection of gravitational waves due to a BNS coalescence \cite{TheLIGOScientific:2017qsa} and GRB170817A \cite{Monitor:2017mdv} have confirmed BNS mergers as a progenitor of short GRBs. 
Targeted searches have been set to look for gravitational-wave transients associated with short and long GRBs, using both modelled and unmodelled search methods \cite{Abbott:2020yvp}. In particular, X-Pipeline \cite{Xpipeline_Sutton_2010} is pipeline based on unmodelled approach that looks  for  excess  power coherent across the network and for this search requires consistency with the sky localization and time window for each GRB.

Core-collapse supernovae (CCSNe) are multimessenger emitters, expected to be a source of EM signals, low energy neutrinos, and gravitational waves which are theoretically not well modelled. Targeted searches for CCSNe detected by astronomical observations \cite{Abbott:2016tdt, Abbott:2019pxc}, have been performed employing the cWB pipeline and imposing the sky  location,  the  source  distance,  and  a  time window  for  the  arrival  time  of  the  gravitational-wave signal. 

A specific searches also have been performed to look for the joint emission of gravitational waves and high energy neutrinos (HEN) \cite{Albert:2018jnn}. The multimessenger search algorithm  \cite{Baret_2012} estimates the significance of the candidate starting from the significance of HEN events and gravitational-wave candidates, identified by the short duration cWB search, and their temporal and directional coincidence.

\section{Continuous} \label{sec:continuous}
\subsection{Potential Sources}

For all cases of gravitational-wave emission, including long-lasting (continuous, persistent) gravitational waves, non-negligible time-varying quadrupole moments are needed. In the case of isolated neutron stars, such emission can be triggered by e.g. elastically and/or magnetically driven deformations: mountains on the stellar surface supported by the elastic strain or magnetic field, free precession, or unstable oscillation modes (e.g. r-modes). Potential sources and emission mechanisms are reviewed in Refs. \refcite{Andersson2011,Lasky2015,Riles2017,Sieniawska2019}. According to the ATNF (Australia Telescope National Facility) Pulsar Database \cite{Manchester2005}, more than 2700 pulsars have been observed via electromagnetic radiation. Assuming that a continuous gravitational-wave signal is approximately twice the neutron star spin frequency\cite{Zimmermann1979} (in the case of elastic/magnetic deformations) or close to $4/3$ of the spin frequency\cite{Owen1998} (Newtonian approximation in case of r-modes), around 300 pulsars from the ATNF Database lie within the most sensitive bandwidths of the LIGO and Virgo detectors. According to studies on population synthesis of isolated radio pulsars, e.g. in Ref. \refcite{Lorimer2005}, there should be $\sim 160,000$ isolated neutron stars  in the Galaxy. Such a relatively large population of potential continuous gravitational-wave emitters motivates searching for persistent signals.

Rotating nonaxisymmetric neutron stars possibly emit gravitational waves with strain amplitude\cite{Ostriker1969, Melosh1969, Chau1970, Press1972, Zimmermann1978} that is on the order of $h_0 \sim 10^{-26}$ :
\begin{equation}
    h_0 = 4.2 \times 10^{-26}\left(\frac{\epsilon}{10^{-6}}\right) \left(\frac{P}{10\textrm{ ms}}\right)^{-2} \left(\frac{d}{1\textrm{ kpc}}\right)^{-1},
    \label{eq:cgw_h0}
\end{equation}
where $\epsilon$ denotes the deformation of an object (also called ellipticity), $P$ is the rotational period and $d$ is the distance to a source.

This amplitude is a few orders of magnitude smaller than signals from coalescing binaries. However, the detectability of a signal, determined by the signal-to-noise ratio ($SNR$), is related not only to $h_0$, but also to the observation time $T$ and the sensitivity of a detector $S$. For example, for the coherent searches such relation is:
\begin{equation}
    SNR \sim h_0 \sqrt{\frac{T}{S}}.
    \label{eq:snr}
\end{equation}
For comparison, the GW150914 event lasted in the LIGO detectors for $T \sim 0.2$ s, with an peak amplitude of $h_0 \sim 10^{-21}$, yielding $SNR\sim24$. For continuous sources,
the expected amplitude is smaller, $h_0 \sim 10^{-25}$, but the data are collected for months or even years. Additionally, the $SNR$ increases with the number of operating detectors in the network $N$, as $SNR \sim \sqrt{N}$. All these aspects (observation time, improvement of the detectors' sensitivities, and new instruments in the global network) make continuous gravitational-wave signals serious candidates for the future detections \cite{Brady1998, Jaranowski2000}.

Neutron stars in binary systems, especially in low mass X-ray binaries (LMXBs) \cite{Lyne2006} are also expected to produce persistent gravitational waves. In such systems, mass is transferred between the neutron star and its companion via Roche-lobe overflow. Searches for signals from such sources require not only the inclusion of complex astrophysical processes, but also orbital parameters of the binary system that additionally modulate the signal\cite{Singh2019}.

The most exotic potential sources of the long-lasting continuous gravitational waves are axion clouds bound to black holes \cite{Arvanitaki2010,Arvanitaki2011}. These bosons clouds, in theory, can be formed due to the energy extraction from the rotation of the black hole. Annihilation or energy level transitions in such a cloud may trigger gravitational-wave emission\cite{Arvanitaki2015}. For stellar mass black holes, and boson masses in the range of ($10^{-14} - 10^{-12}$) eV, the signals would have a frequency
in the sensitivity band of terrestrial detectors and amplitudes that are potentially detectable if emitted within the Galaxy or even outside for a particularly favorable system configurations such as high black hole masses and small boson mass. Even in case of non-detection, analysis results could be used to put interesting constraints on the allowed mass range of ultra-light bosons \cite{DAntonio2018, Palomba2019}.

\subsection{General information about the analyses}

Most of the continuous gravitational-wave searches are dedicated to signals from isolated neutron stars. As was shown previously, $SNR$ increases with longer observation time but coherent analysis of the large amounts of data is computationally expensive. The strategy for the search depends on our \textit{a priori} knowledge of the source, robustness of the data analysis method, and available computational resources. In principle, continuous gravitational-wave searches can be divided into three categories: \textit{i}) Targeted searches in which relevant parameters such as sky position, rotational frequency, and its evolution are known from the electromagnetic observations. Such searches are computationally easy to perform, as they only check if a continuous gravitational-wave signal is associated with the known source and the data can be analysed coherently. Narrow-band searches are a slight modification to the targeted searches, in which a small mismatch between known (from electromagnetic observations) and expected (from gravitational-wave data) frequency parameters is allowed. \textit{ii}) Directed searches are dedicated to the sources with known sky position, but unknown frequency and its evolution. Example sources that are targets for directed searches include core-collapse supernovae remnants or sources close to the Galactic Center. The explored (unknown) parameter space usually requires not only frequency and spin-down (first frequency derivative), but also higher frequency derivatives. It is therefore not possible to perform a long coherent directed search. This means that the data has to be partitioned into shorter time segments, the search performed individually for each segment, the the power from all segments combined following consistent frequency evolution. Such methods are so-called semi-coherent search strategies. \textit{iii}) All-sky (blind) searches are the most computationally expensive types, since only minimal assumptions are made. They are dedicated to signals from completely unknown sources. Such searches are less sensitive than targeted and directed ones. They also require well-optimised data analysis tools and large computational infrastructure. However, they may also detect neutron stars that are not emitting electromagnetically. 

An interesting solution for the limited computational resources was proposed by the \texttt{Einstein@Home} project \cite{Abbott:2008uq, Abbott:2009nc, Aasi:2012fw, TheLIGOScientific:2016uns, Abbott:2017pqa}. The \texttt{Einstein@Home} - formerly used by the LIGO Scientific Collaboration and the Virgo Collaboration - is a volunteer distributed computing project, in which a global network of contributors provides the idle time of their computers to perform computationally intensive data analysis.

So far, no continuous gravitational-wave signal has been confirmed in the LIGO-Virgo data. However, targeted\cite{Abbott:2003yq, Abbott:2004ig, Abbott:2007ce, Abadie:2011md,Collaboration:2009rfa, Aasi:2013sia, Aasi:2014jln, Abbott:2017ylp, Abbott:2017tlp, Abbott:2017cvf, Authors:2019ztc, Abbott:2019bed}, directed\cite{Abadie:2010hv, Aasi:2013jya, Aasi:2014qak, Aasi:2014ksa, Aasi:2015ssf, Abbott:2016tvg, TheLIGOScientific:2016xzw, Abbott:2018qee, LIGOScientific:2019gaw}, blind\cite{Abbott:2005pu, Abbott:2006vg, Abbott:2007td, Abbott:2008uq, Abbott:2008rg, Abbott:2009nc, Abadie:2011wj, Aasi:2012fw, Aasi:2013lva, Aasi:2014mtf, Aasi:2015rar, Abbott:2016udd, TheLIGOScientific:2016uns, Abbott:2017pqa, Abbott:2017mnu, Abbott:2018bwn, Pisarski:2019vxw} and binary searches\cite{Abbott:2006vg, Aasi:2014erp, Abbott:2017hbu, Abbott:2017mwl, Abbott:2019uwg} were intensively performed in the past. In principle, by combining Eqs.~\ref{eq:cgw_h0} and \ref{eq:snr} it is possible to set interesting upper limits. It is especially easy for the targeted searches, where the period and distance of the source are known from the electromagnetic observations. For the known sensitivity of the detector(s) and observation time, one can set signal-to-noise ratio threshold, above which signal should be visible in the data and determine astrophysical information about deformation of the star. For example, for the Crab pulsar (J0534+2200a), $95\%$ credible
upper limits on the fiducial ellipticity was set to be $\epsilon \sim 10^{-5}$\cite{Abbott:2020lqk}.

Below we present several the most commonly used and up-to-date algorithms and methods that are used for various types of continuous gravitational-wave searches.

\subsection{Strategies}

\textit{$\mathcal{F}$-statistic} \cite{Jaranowski1998} is a matched-filtering method, obtained by maximising the likelihood function (function that measures the goodness of fit of a statistical model to a sample of data) with respect to four unknown parameters of the simple emission model (rotating triaxial ellipsoid): strain amplitude $h_0$, wave polarisation angle, initial phase, and the angle between the object's rotational axis and the line of sight (which are henceforth called the extrinsic parameters). This leaves a function of only four remaining parameters: frequency, spin-down and two sky coordinates (called the intrinsic parameters). Thus the dimension of the parameter space that we need to search decreases from 8 to 4. $\mathcal{F}$-statistic is the universal method for all types of searches: targeted, directed and all-sky, as some or all intrinsic parameters can be set as known or be treated as a free parameters. In the case of the blind searches, to improve method efficiency, the $\mathcal{F}$-statistic can be evaluated on the 4-dimensional optimal grid\cite{Pisarski2015} of the intrinsic parameters. According to Eq.~\ref{eq:snr}, detectability of the signal increases with observation time. However, longer timeseries require more computational resources. The solution to this challenge is a hierarchical semi-coherent method, in which data is broken into short segments. In the first stage, each segment is analysed with the $\mathcal{F}$-statistic method. In second stage, the short time segment results are combined incoherently using a certain algorithm (coincidences among candidates\cite{Abbott:2006vg, Abbott:2008uq}, stack-slide method\cite{Brady1998, Brady2000, Cutler2005}, Power-Flux method\cite{Abbott:2007td, Abbott:2008rg, Dergachev2019}, global correlation coordinate\cite{Pletsch2008, Pletsch2009}, Weave method\cite{Wette2018, Walsh2019}). The final follow-up stage is the optimisation procedure\cite{Astone2010, Shaltev2013, Sieniawska2019b} that precisely estimate a global maximum of the $\mathcal{F}$-statistic in the four-dimensional parameter space. An extension of the $\mathcal{F}$-statistic, called $\mathcal{C}$-statistic\cite{Messenger2007}, is dedicated to the signals from LMXBs (neutron stars in binary systems). The strategy is divided into three stages: 1) a wide bandwidth, $\mathcal{F}$-statistic search demodulated for sky position; 2) signal candidates are identified through the frequency-domain convolution of the $\mathcal{F}$-statistic (\textit{a priori} knowledge about the orbital period and orbital semi-major axis is required to construct a template); 3) fully coherent Monte Carlo Markov Chain to follow-up the promising candidates. A similar strategy for binary systems, that uses frequency-domain matched filter, is a $\mathcal{J}$-statistic\cite{Suvorova2016}. In contrast to the $\mathcal{C}$-statistic, this method uses a Hidden Markov Model technique to efficiently track the signal with a wandering frequency on the follow-up stage of the search.

\textit{The Hough transform} and \textit{5n-vector} is a feature extraction technique used in image analysis. In the case of the Frequency-Hough transform \cite{Antonucci2008,Astone2014} the inputs are the most significant time-frequency peaks found in the spectrum and selected above a given threshold. The transform converts the selected time-frequency peaks (peakmap) into lines of the parameter space, identified by the intrinsic frequency and its first time derivative. This method is typically used in the semi-coherent searches i.e., directed and all-sky searches, for which the rotational parameters of the star are unknown. Another type of the Hough transform, used in the continuous gravitational-wave searches, is a Sky-Hough \cite{Krishnan2004}, where peakmaps are generated on the celestial sphere. For searches with known (targeted) or roughly estimated (narrow-band) phase evolution, fully coherent searches can be performed. In such cases it is more convenient to used matched-filtering
techniques like the 5n-vector method\cite{Astone2010a,Astone2012}. This method exploits the typical 5-frequency feature observed in the spectrum of an impinging persistent signal at the detector, when some frequency modulations can be eliminated. These frequency modulations (i.e., Doppler, spin-down plus other relativistic effects) can be removed with a resampling technique or heterodyning the data, using the Band Sampled Data framework\cite{Piccinni2019}. Generalised application of the Hough transform was introduced in Ref. \refcite{Miller2018}. While most of the searches assume some specific process of the gravitational-wave radiation (purely quadrupolar radiation or spin-down due to the oscillations), Generalised Hough models assume the frequency evolution as a power-law, which essentially allows for the analysis of all spin-down orders, in contrast to the Hough searches that only consider first frequency derivatives. This method is best used for rapidly spinning-down neutron stars, i.e. remnants of mergers, such as GW170817 \cite{Abbott:2018hgk} or supernovae remnants. Recently, the Hough method was used for an all-sky search for continuous gravitational-wave signals from unknown neutron stars in binary systems\cite{Covas2020}. The pattern recognition algorithm was applied to every point on the sky. The method replaces the search over the spin-down parameter of isolated sources for the three binary orbital parameters characterizing different possible circular orbits. Machine learning techniques (clustering, Monte Carlo Markov Chains) are used to increase the robustness of the search.

\textit{Bayesian approach}\cite{Dupuis2005, Pitkin2017} takes measurements of the spin, its evolution (up to the second frequency derivative), and sky localisation from the electromagnetic observations. It is typically used (due to the relatively large computational cost) for the targeted searches and application to the directed searches has been proposed\cite{Ming2016,Ming2018}. Unknown parameters of that search are: strain amplitude $h_0$, wave polarisation angle, initial phase and the angle between object's rotational axis, and the line of sight. Due to the Earth’s rotation, the amplitude of the signal recorded by an interferometric detector is time-varying as the source moves through the antenna pattern. The data is demodulated (effects of the Earth's movement and rotation is removed) and binned to short (e.g., 1 min) samples. Then the classical Bayesian approach is adopted. Posterior probability $p(\mathbf{a}|{B_k}, I)$ of the pulsar parameters $\mathbf{a}$ given the binned data, ${B_k}$, is calculated as follows:
\begin{equation}
  p(\mathbf{a}|{B_k}, I) = \frac{p(\mathbf{a}|I) p({B_k}|\mathbf{a}, I)}{p({B_k}| I)},  
\end{equation}
where $\mathbf{a}$ is the set of parameters inferred from data ${B_k}$ given our model $I$, and with likelihood $p({B_k}|\mathbf{a}, I)$. Finally, the parameter estimation is done by numerical marginalisation. Robust and ordinarily used algorithm for the last marginalisation stage is called Markov Chain Monte Carlo\cite{Collaboration:2009rfa, Ashton2018}, in which the parameter space is explored effectively without spending much time in the areas with very low probability densities. Bayesian approach versus maximum-likelihood statistics is summarised in \refcite{Prix2009}.

\textit{TwoSpect Algorithm}\cite{Goetz2011, Aasi:2014erp} is dedicated to the all-sky searches for continuous gravitational waves from unknown sources in binary systems. Similar to the other continuous all-sky searches, the data is divided into short timeseries. The Fourier transform is performed for each data chunk and the power of each Fourier coefficient is computed. Next, each Fourier transform is weighted according to
the noise level and by the antenna pattern of the detector. Time spectrograms over a narrow frequency band are created and the effects of the Earth’s motion are removed. For each spectrogram one has to compute the Fourier transform for powers of each frequency bin  as a function of time and then, using the calculated Fourier coefficients, determine the power spectra of the second Fourier transform. Interesting regions of parameter space are selected if a specific threshold value is exceeded. At this stage, candidates from the doubly-Fourier-transformed data do not depend on any signal template. Finally, the interesting regions are compared with the model templates in order to confirm or reject specific outliers.

\subsection{Future prospects}
Continuous gravitational-wave data analysis is an evolving field. Persisting improvements of the existing methods and software, development of the new strategies, as well as the upgrade of the detectors increase our chances to detect gravitational-waves from isolated neutron stars and ones in LMXBs. A trend that is currently observed, not only in gravitational-wave data analysis but also in many sectors inside and outside science, is the application of machine learning techniques\cite{Miller2019}. Some of them, like clustering or Monte Carlo Markov Chains, were mentioned in the previous subsection. As was pointed out in \refcite{Miller2018}, the Generalised Hough method can be even more universal (including recognition of the different constant or time-varying power laws of the gravitational-wave emission) by using neural networks and random forests. Exploitation of Deep Neural Networks have been proposed as a novel search method for continuous gravitational waves from unknown spinning neutron stars\cite{Dreissigacker2019}, giving promising results on artificial data. Convolutional neural networks can be used not only to detect persistent signal in the detectors' data, but also to determine if a certain signal has instrumental or astrophysical origin\cite{Bayley2020, Morawski2020}. Deep learning can replace traditional methods and reduce the number of promising candidates (as well as computational cost) in the large projects like \texttt{Einstein@Home}\cite{Behnke2015, Papa2016, Singh2017}.

\section{Stochastic} \label{sec:stochastic}

The gravitational-wave background (GWB) is a superposition of signals from astrophysical and cosmological gravitational-wave sources that can be described statistically \cite{Cornish:2015pda}. There are many potential sources, ranging from physics that takes place in the very early Universe such as inflation, phase transitions, and   cosmic strings; to astrophysical sources such as individual rotating neutron stars, supernovae, and the mergers of compact binaries; see Refs. \refcite{Regimbau_2011,stoch_cosmo_review_2018} for reviews of GWB sources. The observable used in searches for a GWB is gravitational-wave energy density per unit logarithmic frequency:
\begin{equation}
    \Omega_{\rm GW}(f) = \frac{f}{\rho_c} \frac{{\rm d} \rho_{\rm GW}}{{\rm d} f}
\end{equation}
where $\rho_c = 3 H_0^2 c^2 / (8\pi G)$ is the energy density needed for a spatially flat Universe, and ${\rm d} \rho_{\rm GW}$ is the gravitational-wave energy density contained in the frequency interval $f$ to $f+{ d}f$.
The standard search uses cross-correlation methods to look for these sources. For an isotropic background, perhaps the most promising source is from compact binaries \cite{TheLIGOScientific:2016wyq,Abbott:2017xzg}. Using a fiducial model and observations from O3a, the median and 90\% credible interval predicted for the GWB from binary black holes and binary neutron stars is $7.2^{+3.3}_{-2.3} \times 10^{-10}$ at 25 Hz, while the 95\% upper limit from O3 for a power-law GWB with a spectral index of $2/3$, consistent with expectations of a GWB from inspiralling binaries, is $3.4 \times 10^{-9}$ at 25 Hz \cite{stoch_O3}. At the design sensitivity of Advanced LIGO and Virgo, we will start to approach the expected level of the background, and a network operating with A+ design sensitivity is sensitive to a significant fraction of the expected range. For detailed reviews of search methods for GWBs, see Refs. \refcite{Allen_Romano_1999,RomanoCornish}, and for a recent review of the state of the field see Ref. \refcite{Christensen_2018}. With each data set, LIGO and Virgo regularly produce search results for isotropic backgrounds  \cite{TheLIGOScientific:2016dpb,LIGOScientific:2019vic,stoch_O3}, cosmic strings \cite{Abbott:2017mem,O3_cosmic_strings}, anisotropic backgrounds \cite{TheLIGOScientific:2016xzw,LIGOScientific:2019gaw}, and scalar and vector polarizations not present in general relativity\cite{Abbott:2018utx}. Contaldi and Renzini\cite{Renzini:2018nee,Renzini:2018vkx,Renzini:2019vmt} have also constrained isotropic and anisotropic backgrounds, using open data from O1 and O2  \cite{Abbott:2019ebz}.

Extensions of these methods have enabled searches for anisotropies in the background\cite{rad_stoch_1,rad_stoch_2,sph_stoch}
\begin{equation}
\Omega_{\rm GW}(f,\Theta) = \frac{f}{\rho_c} \frac{d^3 \rho_{\rm GW}}{d f d^2\Theta}.
\end{equation}
where $\Theta$ refers to the sky direction (angle). Additionally, the radiometer algorithm, built on similar principles, can search for unmodelled signals from persistent narrowband sources. For example, the radiometer can search for gravitational waves from an individual rotating neutron star with accretion, which might be difficult to model analytically.

\subsection{Cross-correlation search}

We can write output timeseries $d_I(t)$ of a set of $N$ detectors as
\begin{equation}
    d_I(t) = h_I(t) + n_I(t)
\end{equation}
where $I=\{1,2,\cdots N\}$ labels the detector, $h_I(t)$ is the signal in detector $I$, and $n_I(t)$ is noise in detector $I$. The fundamental idea of the cross correlation search is that by cross correlating the data over a long period of time, uncorrelated noise in widely separated detectors will tend to cancel, and what remains will be a correlated gravitational wave signal. Denoting the short Fourier transform of the data in a segment of duration $T$ as $\tilde{d_I}(f,t)$, the expectation value of the cross correlation $C(f,t)$ is:
\begin{eqnarray}
    \langle C_{IJ}(f,t) \rangle &\equiv& \frac{2}{T} \langle \tilde{d}^\star_I(f) \tilde{d}_J(f,t) \rangle \nonumber \\
    &=& \frac{2}{T}  \left( \langle \tilde{h}^\star_I(f,t) \tilde{h}_J(f,t) \rangle 
     +  \cancel{\langle \tilde{h}^\star_I(f,t) \tilde{n}_J(f,t) \rangle 
     +  \langle \tilde{n}^\star_I(f,t) \tilde{h}_J(f,t) \rangle }
     +  \cancel{ \langle \tilde{n}^\star_I(f,t) \tilde{n}_J(f,t) \rangle }  \right) \nonumber \\
     &\propto& \Omega_{\rm GW}(f)
\end{eqnarray}
where the second two terms cancel because the gravitational-wave signal and noise are uncorrelated, while the last term cancels if the noise is uncorrelated between the two detectors (we will revisit this assumption later in this section). To optimize the search in real analysis applications, one can introduce a time-and-frequency-dependent filter function $Q_{IJ}(f,t)$ in the cross-correlation by considering the combination $\langle Q_{IJ}(f,t) C_{IJ}(f,t)\rangle$. The best choice of a filter function $Q_{IJ}(f,t)$ depends on the type of GWB signal being studied. The stochastic data analysis group in LIGO-Virgo uses a MATLAB \cite{MATLAB:2020} implementation of the searches below, available at Ref. \refcite{x_stochastic_public}.

\subsubsection{Isotropic GWBs}
For a gravitational-wave background that is \emph{isotropic}, \emph{Gaussian}, \emph{stationary}, and \emph{unpolarized}, the expected GWB signal is independent of time. One can derive an \emph{optimal statistic}\cite{Allen_Romano_1999}:
\begin{equation}
    \hat{C}_{IJ}(f) = \frac{2}{T} \frac{{\rm Re}[\tilde{d}^\star_I(f) \tilde{d}_J(f)]}{\gamma_{IJ} S_0(f)},
\end{equation}
where $\gamma_{IJ}(f)$ is the \emph{overlap-reduction function}\cite{christensen92,Allen_Romano_1999} between the two detectors $I$ and $J$, which accounts for the individual detector response functions, as well as the relative orientation and separation of the detectors. We have also defined the function $S_0(f)\equiv (3H_0^2)/(10\pi f^3)$. This statistic is an unbiased estimator in the sense that $\langle \hat{C}_{IJ}(f) \rangle = \Omega_{\rm GW}(f)$.
In the weak-signal approximation, the variance of this statistic can be estimated as:
\begin{equation}
    \sigma^2_{IJ}(f) \approx \frac{1}{2 T \Delta f}\frac{P_I(f) P_J(f)}{\gamma_{IJ}^2(f) S_0^2(f)}.
\end{equation}
where $P_{I,J}(f)$ are the power spectral densities of the detectors data $d_{I,J}(t)$. One can combine the estimators over different baselines by optimally weighting according to the inverse variance of the baseline, $C(f)=\sum_{IJ} C_{IJ}(f) \lambda_{IJ}(f)$, with the optimal weights given by $\lambda_{IJ}(f)=\sigma^{-2}_{IJ}(f)/\sum_{IJ} \sigma^{-2}_{IJ}(f)$. For a known spectral shape, one can also derive an optimal search statistic.

\subsubsection{Anisotropic GWBs}
For \emph{anisotropic} backgrounds, the strength of the GWB measured by the detectors will change due to changes in the detector response function as Earth rotates through a sidereal day. We assume that the GWB can be factorized as $\Omega_{\rm GW}(f,\Theta)=H(f)P(\Theta)$. Following a procedure very similar to the isotropic case, one is led to an optimal estimate \cite{sph_stoch}:
\begin{equation}
    P_{\alpha} = \sum_\beta [\Gamma^{-1}]_{\alpha \beta} X_\beta
\end{equation}
where $\alpha,\beta$ are indices that depend on the angular coordinates. The indices can label pixels on the sky, or spherical harmonic multipoles $\ell,m$. In this expression, the $P_\alpha$ is the so-called clean map and is an unbiased estimator of the gravitational-wave power, $H(f) \langle P_\alpha \rangle = \Omega_{\rm GW}(f,\alpha)$, and the dirty map $X_{\alpha}$ and Fisher matrix $\Gamma_{\alpha \beta}$ are given by:
\begin{eqnarray}
\label{eq:directional-stochastic-statistics}
X^{IJ}_\beta &=& \sum_t \sum_f \gamma^\star_{IJ,\beta}(f,t) \frac{H(f)}{P_I(f,t) P_J(f,t)} C_{IJ}(f,t) \nonumber \\
\Gamma^{IJ}_{\alpha\beta} &=&\sum_t \sum_f \gamma^\star_{IJ,\alpha}(f,t) \frac{H(f)}{P_I(f,t) P_J(f,t)} \gamma_{IJ,\beta}(f,t)
\end{eqnarray}
The overlap functions $\gamma_\alpha(f,t)$ depend both on the angular coordinates as well as time and frequency \cite{sph_stoch}. 

 The clean map depends on the inverse of the Fisher matrix $\Gamma_{\alpha \beta}$, however the Fisher matrix is singular due to the diffraction limit and due to blind spots in the detector network. Depending on the signal model, there are different approaches for regularizing and inverting the Fisher matrix have been proposed. For \emph{extended} sources on the sky, one can use the \emph{spherical-harmonic decomposition}, and remove multipoles above a certain maximum multipole number $\ell_{\rm max}$. For point-like sources, one assumes that a source can only be in one pixel and then use only the diagonal elements of the Fisher matrix, so $\Gamma^{-1}=\Gamma_{\Theta \Theta}^{-1}$. This is known as the \emph{radiometer algorithm} \cite{rad_stoch_1}. The inverse of the Fisher matrix is also used as an estimator of the variance of the clean map.   

One can also consider \emph{narrow-band} anisotropic gravitational-wave sources. For this search, one looks at the sky map as a function of frequency. Historically, due to computational resources, this method has only been applied to a limited number of promising sky directions: the remnant of supernova 1987A, the X-ray binary Scorpius X-1, and the Galactic Center. More efficient algorithms have enabled the radiometer search for all frequencies and sky directions \cite{Thrane:2015aua,Goncharov:2018ufi}.

Neglecting small Doppler shift effects, the expressions for the search statistic in Eq.~\eqref{eq:directional-stochastic-statistics} are invariant under time translations by a sidereal day. This allows one to \emph{fold} the data from an entire observing run to one sidereal day \cite{Ain:2015lea}, by combining data segments which are separated by a sidereal day. Folding reduces the amount of data that needs to be analyzed by a factor of the number of sidereal days in the run, and enables new searches such as an all-sky-all-frequency radiometer \cite{Thrane:2015aua,Goncharov:2018ufi}, as well as efficient algorithms to perform the directional search such as PyStoch \cite{Ain:2018zvo}.

\subsection{Upper limits on isotropic sources}

A useful and straightforward way to constrain backgrounds that can be approximated as a power law is to use a power-law integrated (PI) curve. The PI curve is a function of frequency. For each frequency, the value of the PI curve is given by the amplitude of the largest power-law GWB at that frequency that can be detected with a given signal-to-noise ratio, considering all spectral indices \cite{locus}. The PI curve naturally takes into account the gain from integrating over different frequency bins. PI curves are typically released alongside the LIGO-Virgo analysis papers; as of this writing, the most recent PI curves come from the O3 analysis and are available at Ref.  \refcite{stoch_o3_data_release}.

We can also perform a more detailed analysis that can be used for inference. Given a parameterized model for the GWB, $\Omega_{M}(f;\theta_M)$ for some parameters $\theta_M$, we can perform Bayesian parameter estimation on isotropic GWBs by using the Gaussian likelihood\cite{StochPE}:
\begin{equation}
    p(\hat{C}_{IJ}(f_k)|\theta_M) \propto \exp\left[\sum_{k} -\frac{1}{2} \left( \frac{\hat{C}_{IJ}(f_k) - \Omega_M(f_k;\theta_M)}{\sigma_{IJ}(f_k)}\right)^2 \right].
\end{equation}
Using Bayes' theorem, we construct the posterior by multiplying by a prior for the parameters in the model $\theta_M$, $p(\theta_M|\hat{C}_{IJ}(f_k))\propto p(\hat{C}_{IJ}(f_k)|\theta_M) p(\theta_M)$.
For simple models with a few parameters one can evaluate the posterior on a grid; for more complex models one can use sampling algorithms. Note the input here is the frequentist statistic $\hat{C}_{IJ}(f)$ derived from the cross-correlation search, and not the strain data (as is used in other signal-classes). This hybrid frequentist-Bayesian technique has been shown to be equivalent to a full Bayesian analysis starting from the strain data\cite{Matas:2020roi}, under a set of mild conditions satisfied by LIGO-Virgo searches.

Examples of models that have been used in LIGO analyses include
\begin{itemize}
    \item \emph{Power-law models.} A common choice is to place upper limits on pure power-law models, $\Omega_{\rm GW}(f)=\Omega_{\rm ref}(\frac{f}{f_{\rm ref}})^\alpha$, which is an approximation to many astrophysical and cosmological models in the LIGO-Virgo band.
    \item \emph{Astrophysical or cosmological models} of $\Omega_{\rm GW}(f)$. For example, in O3 the tension of cosmic strings was constrained considering a GWB produced by gravitational-wave emission from cusps, kinks, and kink-kink collisions\cite{O3_cosmic_strings}.
    \item \emph{Vector- and scalar-polarized GWBs}. To perform this search, one relies on the fact that tensor, vector and scalar polarized GWs interact differently with a gravitational-wave detector and lead to different overlap reduction functions\cite{TestingGR_stoch}. Since general relativity predicts only tensor polarized GWs, searching for backgrounds with alternative polarizations is a test of GR.
    \item \emph{Apparent GWB from correlated noise}. Correlated noise, for instance from global magnetic fields that are coherent between different interferometers and couple into the gravitational-wave strain channel, can appear as an apparent GWB. Recently, a Bayesian framework was developed to parameterize the magnetic coupling and simultaneously fit for an apparent magnetic background along with a GWB \cite{Meyers:2020qrb}.
    \item  \emph{Combinations of GWB and other data sets. }Additionally, the gravitational-wave data can be combined with other measurements, for example
    of individual compact binaries. This approach has been used to constrain the merger rate for binary black holes at large redshifts, combining stochastic and CBC measurements\cite{Callister:2020arv,stoch_O3}.
\end{itemize}
This list is not exhaustive. The cross-correlation spectra $\hat{C}_{IJ}(f_k)$ for O3 are publicly available \cite{stoch_o3_data_release} and can be used for Bayesian inference with any GWB model.

\subsection{Upper limits on anisotropic sources}

In the absence of evidence for an anisotropic GWB, we can place upper limits on the GWB as a function of sky-direction and frequency. In broad band radiometer analyses, for different gravitational-wave signal models $H(f)$, we place upper limits on the GWB as a function of sky direction $\Theta$. In a narrow band radiometer analysis we place upper limits on the gravitational-wave strain from specific astrophysical targets on the sky such as supernova 1987A, Scorpius X-1 and the Galactic Center. In the spherical harmonic analysis we place upper limits on the angular gravitational-wave power $[C_\ell]^{1/2}$ in different spherical harmonic modes $\ell$. This is similar to the estimators of anisotropy in cosmic microwave background analyses. The square of angular gravitational-wave power $C_\ell$ is obtained from the clean map $P_{\ell m}$ using the expression:
\begin{equation}
    \hat C_{\ell} = \left(\frac{2\pi^2 f_{\rm ref}^3}{3H_0^2}\right)^2\frac{1}{1 + 2 \ell} \sum_{m=-\ell}^{\ell} \left[|\hat{P}_{\ell m}|^2 - (\Gamma^{-1})_{\ell m,\ell m}\right]
\end{equation}
where the last term in the above expression prevents a bias in the estimator the $\hat{C}_{\ell}$'s \cite{sph_stoch}.
In addition to $C_\ell$, in the spherical harmonics analysis we also produce upper limits on the GWB as function of sky direction applicable to extended sources on the sky. For example, the upper limits on gravitational-wave sky maps and $C_\ell$ using LIGO-Virgo O2 run data are available in Ref.~\refcite{LIGOScientific:2019gaw}. The upper limits on $C_\ell$ from O2 run are more than an order magnitude above the theoretical predictions \cite{Jenkins:2018uac,Cusin:2018rsq}. However using next generation gravitational-wave detectors with significantly improved sensitivity, we might be able to measure the anisotropies in GWB.  

\subsection{Other approaches}

Other approaches have been suggested that may be more sensitive to certain sources of the stochastic background. Drasco and Flanagan in 2002 suggested that for intermittent GWBs, such as a popcorn-like signal from binary black holes, a search statistic which builds in a probability for a signal to be present in only a fraction of the data may detect a GWB signal sooner, at the price of introducing additional computational cost\cite{Drasco:2002yd}. This idea has been explored in subsequent work\cite{popcorn1,popcorn2,popcorn3}, although not yet applied to real data. Smith and Thrane in 2018 proposed an implementation of this idea targeting the binary black hole background using a modeled approach for individual BBHs\cite{tbs_methods}. 

Correlated noise is a potential issue for all cross-correlation searches. Schumann resonances\cite{Schumann_theory} can generate correlated magnetic fields that can couple into interferometers\cite{Schumann_1}. In addition to the Bayesian approach explained here, studies have proposed the use of Wiener filtering (linear subtraction) to remove magnetic noise\cite{Schumann_2,Schumann_3,Schumann_4}. Another approach is \emph{gravitational-wave geodesy} \cite{Callister:2018ogx}. Inspired by the idea of sky-scrambles in pulsar timing array data analysis\cite{Taylor:2016gpq}, this approach checks for consistency between the recovered and expected overlap reduction function.

The radiometer can be extended to look for narrowband point sources at all frequencies and in all sky directions\cite{Thrane:2015aua,Goncharov:2018ufi}. Finally, searches for dark photon dark matter can be developed by extending the isotropic search to look for narrowband signals\cite{Pierce:2018xmy}.

In future ground-based detectors, studies have shown that an astrophysical foreground can be subtracted leaving a residual background of $\sim 3\times 10^{-12}$ at 10 Hz for a network of 5 third generation detectors \cite{Regimbau:2016ike,Sachdev:2020bkk}. For alternative approaches to this problem, see\cite{Biscoveanu:2020gds,Sharma:2020btq}.

\section*{Acknowledgments}

This article has been assigned document number LIGO-P2100066 and VIR-0297A-21.  

The authors wish to personally thank Marie-Anne Bizouard, Giancarlo Cella, Ik Siong Heng, Ian Jones, Fiona Panther, Laura Nuttall, Francesco Pannarale, Joseph Romano, John Whelan, and Gramham Woan for their thoughtful review of this work and comments which have greatly improved it.

This material is based upon work supported by NSF’s LIGO Laboratory which is a major facility
fully funded by the National Science Foundation.
The authors also gratefully acknowledge the support of
the Science and Technology Facilities Council (STFC) of the
United Kingdom, the Max-Planck-Society (MPS), and the State of
Niedersachsen/Germany for support of the construction of Advanced LIGO 
and construction and operation of the GEO600 detector. 
Additional support for Advanced LIGO was provided by the Australian Research Council.
The authors gratefully acknowledge the Italian Istituto Nazionale di Fisica Nucleare (INFN),  
the French Centre National de la Recherche Scientifique (CNRS) and
the Netherlands Organization for Scientific Research, 
for the construction and operation of the Virgo detector
and the creation and support  of the EGO consortium. 
The authors also gratefully acknowledge research support from these agencies as well as by 
the Council of Scientific and Industrial Research of India, 
the Department of Science and Technology, India,
the Science \& Engineering Research Board (SERB), India,
the Ministry of Human Resource Development, India,
the Spanish Agencia Estatal de Investigaci\'on,
the Vicepresid\`encia i Conselleria d'Innovaci\'o, Recerca i Turisme and the Conselleria d'Educaci\'o i Universitat del Govern de les Illes Balears,
the Conselleria d'Innovaci\'o, Universitats, Ci\`encia i Societat Digital de la Generalitat Valenciana and
the CERCA Programme Generalitat de Catalunya, Spain,
the National Science Centre of Poland and the Foundation for Polish Science (FNP),
the Swiss National Science Foundation (SNSF),
the Russian Foundation for Basic Research, 
the Russian Science Foundation,
the European Commission,
the European Regional Development Funds (ERDF),
the Royal Society, 
the Scottish Funding Council, 
the Scottish Universities Physics Alliance, 
the Hungarian Scientific Research Fund (OTKA),
the French Lyon Institute of Origins (LIO),
the Belgian Fonds de la Recherche Scientifique (FRS-FNRS), 
Actions de Recherche Concertées (ARC) and
Fonds Wetenschappelijk Onderzoek – Vlaanderen (FWO), Belgium,
the Paris \^{I}le-de-France Region, 
the National Research, Development and Innovation Office Hungary (NKFIH), 
the National Research Foundation of Korea,
the Natural Science and Engineering Research Council Canada,
Canadian Foundation for Innovation (CFI),
the Brazilian Ministry of Science, Technology, and Innovations,
the International Center for Theoretical Physics South American Institute for Fundamental Research (ICTP-SAIFR), 
the Research Grants Council of Hong Kong,
the National Natural Science Foundation of China (NSFC),
the Leverhulme Trust, 
the Research Corporation, 
the Ministry of Science and Technology (MOST), Taiwan,
the United States Department of Energy,
and
the Kavli Foundation.
The authors gratefully acknowledge the support of the NSF, STFC, INFN and CNRS for provision of computational resources.

\section*{References}

\bibliographystyle{ws-mpla}
\bibliography{refs}

\end{document}